\documentclass[prb,twocolumn,superscriptaddress,showpacs,floatfix]{revtex4}
\usepackage{graphicx}
\begin{document}
\newcommand{\RR}{\mathrm{\mathbf{R}}}
\newcommand{\rr}{\mathrm{\mathbf{r}}}
\newcommand{\defin}{\stackrel{def}{=}}

\title{External field control of donor electron exchange at the Si/SiO$_2$ interface}
\author{M.J. Calder\'on}

\affiliation{Condensed Matter Theory Center, Department of Physics,
University of Maryland, College Park, MD 20742-4111}

\author{Belita Koiller}

\affiliation{Instituto de F\'{\i}sica, Universidade Federal do Rio de
Janeiro, Caixa Postal 68528, 21941-972 Rio de Janeiro, Brazil}

\author{S. {Das Sarma}}

\affiliation{Condensed Matter Theory Center, Department of Physics,
University of Maryland, College Park, MD 20742-4111}

\date{\today}

%$\lesssim$

%$\gtrsim$

\begin{abstract}

We analyze several important issues for the single- and two-qubit operations in Si quantum
computer architectures involving P donors close to a SiO$_2$
interface. For a single donor, we investigate the donor-bound electron
manipulation (i.e. 1-qubit operation) between the donor and the interface by electric and
magnetic fields. We establish conditions to keep a donor-bound state at the
interface in the absence of local surface gates, and estimate the
maximum planar density of donors allowed to avoid the formation of a
2-dimensional electron gas at the interface. We also calculate the
times involved in single electron shuttling between the donor and
the interface. For a donor pair, we find that under certain conditions
the exchange coupling (i.e. 2-qubit operation) between the respective
electron pair at the interface may be of the same order of magnitude
as the coupling in GaAs-based two-electron double
quantum dots where coherent spin manipulation and control has been
recently demonstrated (for example for donors $\sim 10$~nm below the
interface and $\sim 40$~nm  apart, $J\sim 10^{-4}$~meV),
opening the perspective for similar experiments to be performed in Si.

\end{abstract}

\maketitle
\section{Introduction}
\label{sec:intro}
Doped Si is a promising candidate for quantum
computing~\cite{Kane} due to its scalability properties, long spin
coherence
times,~\cite{sousa03,tyryshkin03,abe04,witzel05,tyryshkin06,witzel06}
and the astonishing progress on Si technology and miniaturization
in the last few decades (Moore's law). The experimental production
of a working qubit depends on precise positioning (of the order of
\AA)~\cite{KHD1,Kane_MRS} of donors in Si and the quantum control
of the donor electrons by local gates placed over an oxide layer
above the donors. The required accuracy in donor positioning has
not been yet achieved, although there are increasing efforts in
this direction, using top-down techniques, i.e. single ion
implantation (with tens of nm
accuracy),~\cite{schenkel03,shinada05,jamieson05} and bottom up
techniques, i.e. positioning of P donors  on a mono-hydride
surface via STM (with $1$~nm accuracy) with subsequent Si
overgrowth.~\cite{obrien01,schofield03}

\begin{figure}
\begin{center}
\resizebox{70mm}{!}{\includegraphics{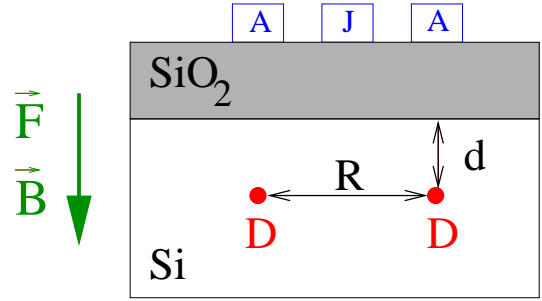}}
\caption{\label{fig:setup}(Color online) Schematic view of Kane's
  quantum computer. Donors D are a distance $d$ from the Si/SiO$_2$
  (001) interface and a distance $R$ from each other. Surface A- and
  J-gates control one and two qubit operations.
  In the present study, we consider uniform electric $F$ and magnetic
  $B$ fields applied in the $z$-direction. }
\end{center}
\end{figure}
In the original doped Si based quantum computer proposal,~\cite{Kane}
illustrated in Fig.~\ref{fig:setup}, the qubits are the donor nuclear spins,
and the hyperfine interaction between these and the donor
electron spins is used to perform single qubit operations
(rotations).
The strength of the hyperfine interaction is manipulated by local
surface gates, the so called A-gates, which move the
electron between the donor and an interface with SiO$_2$. Exchange
between neighboring donors, tuned by surface 'exchange' gates (J-gates),
would control two-qubit operations. Exchange gates were originally
proposed for a double
quantum dot geometry in GaAs.~\cite{Exch} Related proposals in Si use
the electron spin as qubit~\cite{Vrijen,levy01,friesen03} or the electron
charge.~\cite{hollenberg1} Charge coherence in Si is much shorter
($\sim 200$ ns)~\cite{gorman05} than the spin coherence $T_2 \sim 1$
ms, which can be further enhanced by isotopic
purification,~\cite{sousa03,witzel05,tyryshkin06} making spin qubits
in general more attractive than charge qubits for actual
implementations.
On the other hand, direct detection of a single spin is a very
difficult task,~\cite{Rugar04,xiao04,koppens06} while a fraction of a
single electron charge can be easily detected with state-of-the-art
single electron tunneling (SET) devices.
As a result, ingenious spin-to-charge conversion mechanisms,
that would allow the electron spin state to be inferred according to the
absence or presence of charge detected by an SET at the surface, have
been proposed,
e.g.~Refs.~\onlinecite{hollenberg2,greentree05,stegner06,double-donor}.

Doped Si has two main advantages over GaAs quantum dots: (i) The much
longer spin coherence times, that can be enhanced by isotopic
purification (note that all isotopes of Ga and As have nuclear spins
so the spin coherence time in GaAs cannot be improved via isotopic
purification), and (ii) the identical Coulomb potentials created by
donors as opposed to variable quantum dot well shapes produced by
surface gates on a 2-dimensional electron gas (2DEG). Despite this latent
superiority of Si, progress in GaAs has been much faster,
\cite{petta05,koppens05,johnson05,koppens06,laird06} in particular due
to the fact that the electrons, being at the device surface,  are easier to
manipulate and detect.
Another Si handicap is that the exchange between donors in bulk Si
oscillates, changing by orders of magnitude when the relative position
of neighboring donors changes by small distances  $\sim$
\AA.~\cite{KHD1,KHD2}
This is caused by interference effects between the six degenerate
minima in the Si conduction band. However, as discussed below, this
degeneracy is partially lifted at the interface, thus the oscillatory
behavior may not represent such a severe limitation for interface
states as compared to the donor bulk states.

In the following, we analyze the manipulation of donor electrons close
to a Si/SiO$_2$ interface by means of external uniform electric and
magnetic fields.~\cite{QC-PRL06,CKDmag} In
Sec.~\ref{sec:single-donor} we introduce the model for an isolated
donor and discuss the  interface and the donor ground states, which
are calculated variationally. We also analyze the shuttling between
the interface and the donor, including the effect of a magnetic field.
In Sec.~\ref{sec:donor-pair} we study the conditions to avoid the
formation of a 2DEG at the interface, and
discuss the advantages and actual feasibility of performing two-qubit
operations at the interface. A summary and conclusions are given in
Sec.~\ref{sec:discussion}.
\section{Single donor}
\label{sec:single-donor}
\subsection{Model}
\label{subsec:model}
\begin{figure}
\begin{center}
\resizebox{80mm}{!}{\includegraphics{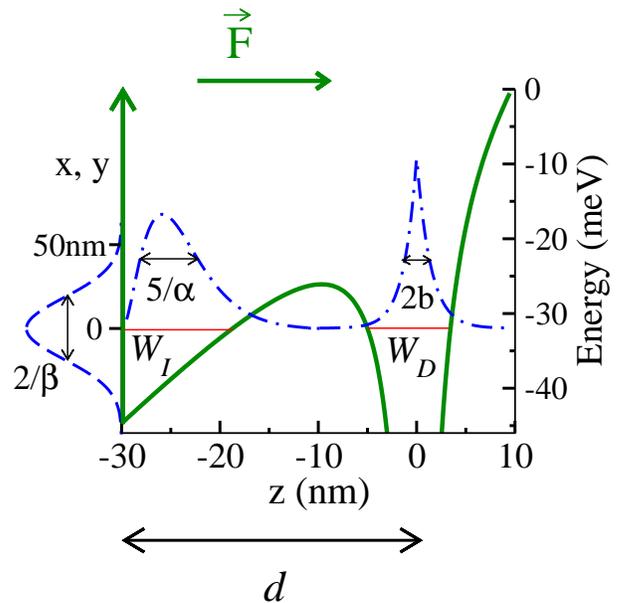}} \caption{(Color
online) Double-well potential formed by the donor nuclear charge
(W$_D$), the applied electric field, and the interface with the
oxide (considered to be $\infty$, i.e., impenetrable). The
interface well W$_I$ consists of the electric field and the
interface in the $z$-direction and also includes the nuclear
charge and its image in the $xy$-plane [(see
Fig.~\ref{fig:energy-rho}(a)].  The dashed lines represent the
decoupled ground eigenstates in each well $\psi_I$ and $\psi_D$.
The thin horizontal lines indicate the expectation value of the
energy in each well. The potential, wave-functions and energies
depicted correspond to $d=30$~nm and $F=13.5$ kV/cm $\approx
F_{\rm c}(30 \,{\rm nm})$.} \label{fig:potentialF}
\end{center}
\end{figure}
We consider initially a single donor a distance $d$ from a Si/SiO$_2$ (001)
interface. As a simple model for the A-gate effects, a
uniform electric field is applied in the $z$-direction, as illustrated in
Fig.~\ref{fig:setup}.
The inter-donor distance $R$ is assumed to be large enough so that
each donor can be treated as an isolated system. The conduction band
of Si has six equivalent minima located along the $\Delta$ lines. As
discussed below, it is reasonable to treat this system within the
single-valley effective mass
approximation\cite{Kohn,stern67,martin78,macmillen84,ando82} leading
to the following Hamiltonian for the donor electron
\begin{equation}
H = T + e F z -{{e^2}\over{\epsilon_{\rm Si}
r}}+{{e^2 Q}\over{\epsilon_{\rm Si}\sqrt{\rho^2+(z+2d)^2}}}-{{e^2
Q}\over{4\epsilon_{\rm Si}(z+d)}}~.
\label{eq:h}
\end{equation}
We also consider an applied magnetic field along $z$: In this case the
vector potential ${\bf A} = B \left(y,-x,0 \right)/2$ is included in
the kinetic energy term,
 $T=\sum_{\eta=x,y,z} \hbar^2/(2 m_\eta) \left[i\partial/\partial  \eta+e
A_\eta/(\hbar c) \right]^2 $. The effective masses in Si are
 $m_x=m_y=m_\perp = 0.191 \,m$, and
 $m_z=m_\|=0.916\,m$.
The second term is the electric field linear potential,
the third is the donor Coulomb potential, and
the last two
 terms (with $\rho^2=x^2+y^2$) take account of the charge images of
 the donor and the electron,  respectively.
 $Q={{(\epsilon_{\rm SiO_{2}}-\epsilon_{\rm Si})}/{(\epsilon_{\rm
 SiO_{2}}+\epsilon_{\rm Si})}}$,
 where $\epsilon_{\rm Si}=11.4$ and  $\epsilon_{\rm SiO_{2}}=3.8$.  In this
 case $Q<0$ and, therefore, the images have the same sign as the
 originating charges.
 In rescaled atomic units, $a^*={{\hbar^2\epsilon_{Si}}/{m_\perp e^2}} =
 3.157$~nm and $Ry^*={{m_\perp e^4}/{2\hbar^2\epsilon_{Si}^2}}=
 19.98$~meV, and the Hamiltonian is written

\begin{eqnarray}
H &=&- {{\partial^2}\over{\partial x^2}} - {{\partial^2}\over{\partial
 y^2}}- \gamma {{\partial^2}\over{\partial
 z^2}}+ {{1}\over{4}} \mu^2 \rho^2 + i \mu (y \partial_x-x\partial_y)
\nonumber \\ &+&\kappa e F z -{{2}\over{r}}+{{2Q}\over{\sqrt{\rho^2+(z+2d)^2}}}-{{Q}\over{2(z+d)}}~,
\label{eq:h-effunits}
\end{eqnarray}
where
$\gamma=m_\perp/ m_\|$, $\mu^2=a^{*4}/\lambda_B^4$ with
$\lambda_B=\sqrt{{\hbar}/eB}$ the magnetic length, $\kappa=3.89 \times 10^{-7} \epsilon_{Si}^3
\left({{m}/{m_\perp}}\right)^2$ cm/kV, and the electric field $F$ is
given in kV/cm.

The system under study consists of a shallow donor, P  in particular,
immersed in Si a distance $d$ from a SiO$_2$ barrier, which we assume
to be infinite (impenetrable).
When no external field is present, the electron is bound to the donor
potential well $W_D$.
When an electric field $F$ is applied in the $z$-direction, a
triangular well $W_I$ is formed next to the interface.
The interface well $W_I$ also includes the donor and its image Coulomb
potentials at the interface ($z=-d$) which, under special
circumstances discussed below, are strong enough to localize the electron
in the $xy$-plane. The interface well $W_I$ and the donor Coulomb
potential $W_D$ form an asymmetric double well, as
illustrated in Fig.~\ref{fig:potentialF}.

The Hamiltonian in Eq.~(\ref{eq:h-effunits}) is solved in the basis
formed by $\psi_D$ and $\psi_I$, which are the ground
eigenstates of each of the decoupled wells $W_D$ and $W_I$.
The Hamiltonian is written
\begin{equation}
H=T+V_I+V_D+{{2}\over{\sqrt{\rho^2+d^2}}}
\label{H}
\end{equation}
where the last term is added to avoid double counting of the impurity
Coulomb potential at the interface included both in $V_I$ and
$V_D$, which are defined as
\begin{equation}
V_I=2{{Q-1}\over{\sqrt{\rho^2+d^2}}} +\kappa eFz-{{Q}\over{2(z+d)}} \,\,,
\label{eq:V_I}
\end{equation}
and
\begin{equation}
V_D=-{{2}\over{r}}\,.
\end{equation}
The first term in Eq.~(\ref{eq:V_I}) is the sum of the donor Coulomb
potential and its image charge potential at the interface.

The Hamiltonian in the non-orthogonal basis $\{\psi_D,\psi_I\}$ reads
\begin{equation}
\left( \begin{array}{cc}
 H_{DD} &  H_{ID} \\
 H_{ID} & H_{II} \end{array} \right)~,
\label{eq:2by2}
\end{equation}
where $H_{AB}=\langle \psi_A | H |\psi_B \rangle$ with $A, B =I,
D$. Diagonalization gives the two eigenstates $\Psi^+$ and
$\Psi^-$ with eigenenergies $E^+$ and $E^-$ which
show anticrossing behavior with a minimum gap
when $H_{DD} = H_{II}$. This point defines the characteristic field
$F_{\rm c}$, illustrated in Fig.~\ref{fig:potentialF}, which is relevant for the tunneling process discussed in
detail in Sec.~\ref{subsec:shuttling}.

\subsection{Interface state $\psi_I$}
\label{subsec:interface}

\begin{figure}
\begin{center}
\resizebox{80mm}{!}{\includegraphics{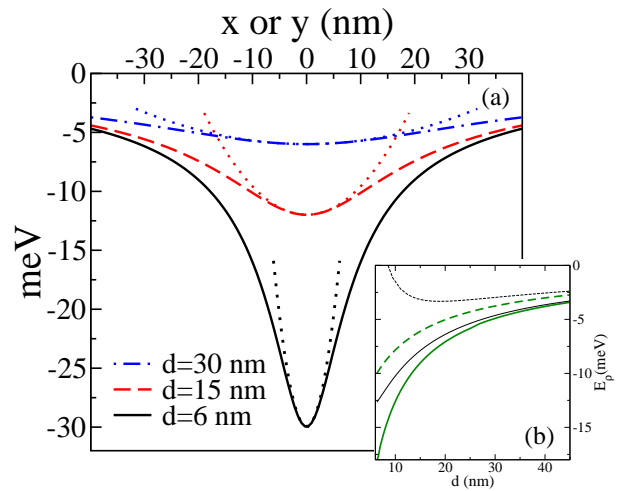}}
\caption{\label{fig:energy-rho}(Color online) (a) Confining
potential at the interface $xy$-plane due to the Coulomb attraction of the
donor nucleus and its image
$V_{I}^{\rho}=2(Q-1)/\sqrt{\rho^2+d^2}$. Dotted lines correspond to
the harmonic approximation of the potential $V_{\rm parab} (\rho)=
(Q-1)(2/d-\rho^2/d^3)$. (b) Ground state energy $E_{\rho}$ (solid
lines) and first excited state energy $E'_{\rho}$ at the interface
(dashed lines). The thick lines correspond to the potential
$V_{I}^{\rho}$ while the thin lines correspond to the parabolic
approximation. As $d$ increases, the solution for the parabolic approximation
approaches the one for the $V_I^\rho$ potential, as expected.
}
\end{center}
\end{figure}

It is convenient to write the interface potential $V_I$ as a sum of
purely $z$- and purely $\rho$-dependent terms:
\begin{eqnarray}
V_I&=&V_I^z + V_I^{\rho} \,,\\
V_I^z&=&\kappa eFz-{{Q}\over{2(z+d)}} \,,\\
V_I^{\rho}&=&2{{Q-1}\over{\sqrt{\rho^2+d^2}}} \,.
\end{eqnarray}
The $V_I^z$ component is the triangular well plus the electron image charge
potential, while the $V_I^{\rho}$ component is the sum of the impurity and its image
potential at the interface. $V_I^{\rho}$ is plotted in
Fig.~\ref{fig:energy-rho}(a) for three different values of $d$.
Curves corresponding to the parabolic approximation of the potential,
\begin{equation}
V_{\rm parab} (\rho)= (Q-1)\left({2 \over d}-{\rho^2 \over d^3}\right) \,  ,
\label{eq:v-rho-parab}
\end{equation}
are also shown, and it is clear that the harmonic approximation works better for the
larger distances $d$.
Electron confinement at the interface $xy$-plane is provided by
$V_I^{\rho}$, while the uniform electric field and the oxide confine
the electron along the $z$-direction.
For certain values of $d$ and $R$, $V_I^{\rho}$ is deep enough to
localize the individual donor electrons (with no need of local A-gates) and keep them from forming a
2DEG at the interface: The necessary
conditions are discussed in Sec.~\ref{subsec:planar-density}. The
spacial localization of the electrons at the interface is a necessary
condition for Si-based quantum computing if qubit read-out
takes place at the interface.~\cite{Kane}

\begin{figure}
\begin{center}
\resizebox{80mm}{!}{\includegraphics{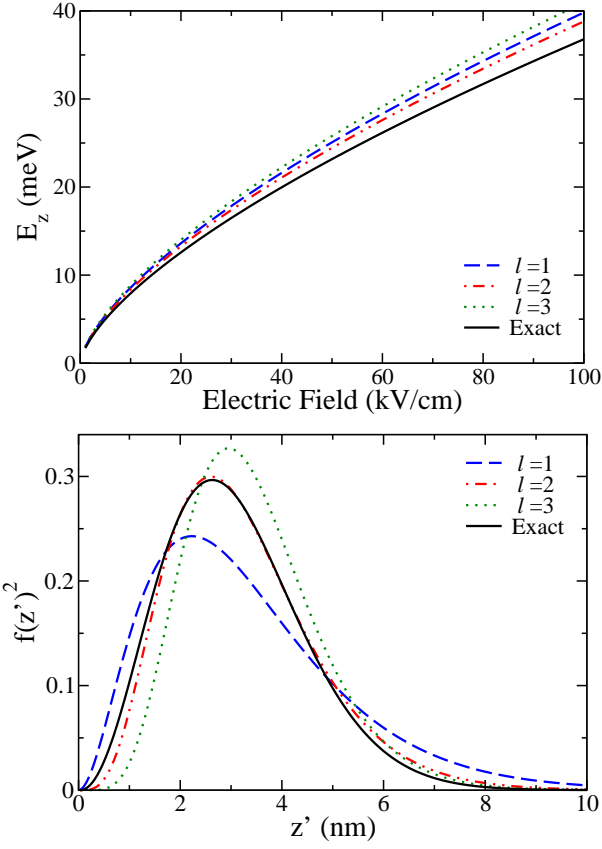}}
\caption{\label{fig:barrierz}(Color online) (a) $E_z$ for different
trial functions $f(z)$ compared with the exact solution of the
infinite triangular potential.~\cite{stern72}(b) Resulting
wave-function for $F=50$ kV/cm. $z'=z+d$. Both the energy and wave-function for
$\ell=2$ give the best match to the exact solution.}
\end{center}
\end{figure}

The ground state at the interface is calculated by solving $H_I=T+V_I$
variationally with a separable trial function
\begin{equation}
\psi_I=f(z) g(\rho) \,\, .
\label{eq:psiI}
\end{equation}
For the $z$-part we use
\begin{equation}
f(z)={{\alpha^{(2\ell +1)/2}}\over{\sqrt{(2 \ell)!}}}
 (z+d)^{\ell} e^{-\alpha (z+d)/2} \,\,\,
\end{equation}
for $z > -d$.
The infinite barrier at the interface is taken into account
by forcing the ground state to be zero at the interface, so
$f(z)=0$ for $z \le -d$. $\alpha$ is a variational
 parameter that minimizes the contribution to the energy $\langle f(z) |T^z+ V_I^{z} |
 f(z) \rangle $. The most suitable value for $\ell$ is chosen by
 comparing $E_z=\langle f(z) |T^z+ \kappa eF z | f(z) \rangle$ and
 $f(z)$
 with the exact solution of an infinite triangular well~\cite{stern72}
\begin{equation}
f^{\rm exact}(z')= Ai\left(\sqrt[3]{2 m_z e F \over \hbar^2}
  \left[z'-{E_0 \over eF} \right]\right) \, ,
\end{equation}
where $Ai$ is the Airy function, $z'=z+d$, and $E_0$ is the ground state energy
\begin{equation}
E_0 =E_z^{\rm exact}=\sqrt[3]{{\hbar^2\over 2 m_z} \left(0.7587 \, \pi e F \right)^{2}} \, .
\label{eq:Etriangle}
\end{equation}
The results are shown in Fig.~\ref{fig:barrierz} for $\ell=1$, $2$, and $3$.
$\ell=2$ gives the best agreement with the exact solution for both the
energy and the wave-function and is therefore adopted in what follows.

For the $\rho$-part we use the ansatz
\begin{equation}
g(\rho)= {{\beta}\over{\sqrt{\pi}}} e^{-\beta^2
  \rho^2/2} \,,
\end{equation}
with the variational parameter $\beta$ calculated by minimizing
$E_{\rho}=\langle g(\rho) |T^{\rho}+ V_I^{\rho} | g(\rho) \rangle
$. We have checked that this gaussian form gives lower energy than
an exponential  $e^{-\eta \rho/2}$ (as a reminiscent of the donor
wave-function) for distances $d >1$~nm.

In Fig.~\ref{fig:energy-rho}(b) we plot the energy $E_{\rho}$ for the
ground state and the first excited state ($g'(\rho) \propto x
g_{\beta'}(\rho)$) for both the variational solution adopted here and the parabolic
approximation of $V_I^{\rho}$. The parabolic approximation gives an
underestimation of the binding energies and diverges at short
distances $d$ (not shown).  To guarantee that the electron remains
bound, and at the ground state, the operating temperature has to be
lower than the energy difference between the ground and the first
excited states
$k_B T \ll min(|E_{\rho}|,E'_{\rho}-E_{\rho})$. For $d=30$~nm, the
excitation gap $E'_{\rho}-E_{\rho} \sim 1$~meV. This limits the
operating temperature to a few Kelvin (in current experiments,
temperatures as low as $0.1$ K are being
used).~\cite{ferguson06,kenton06}

\begin{figure}
\begin{center}
\resizebox{60mm}{!}{\includegraphics{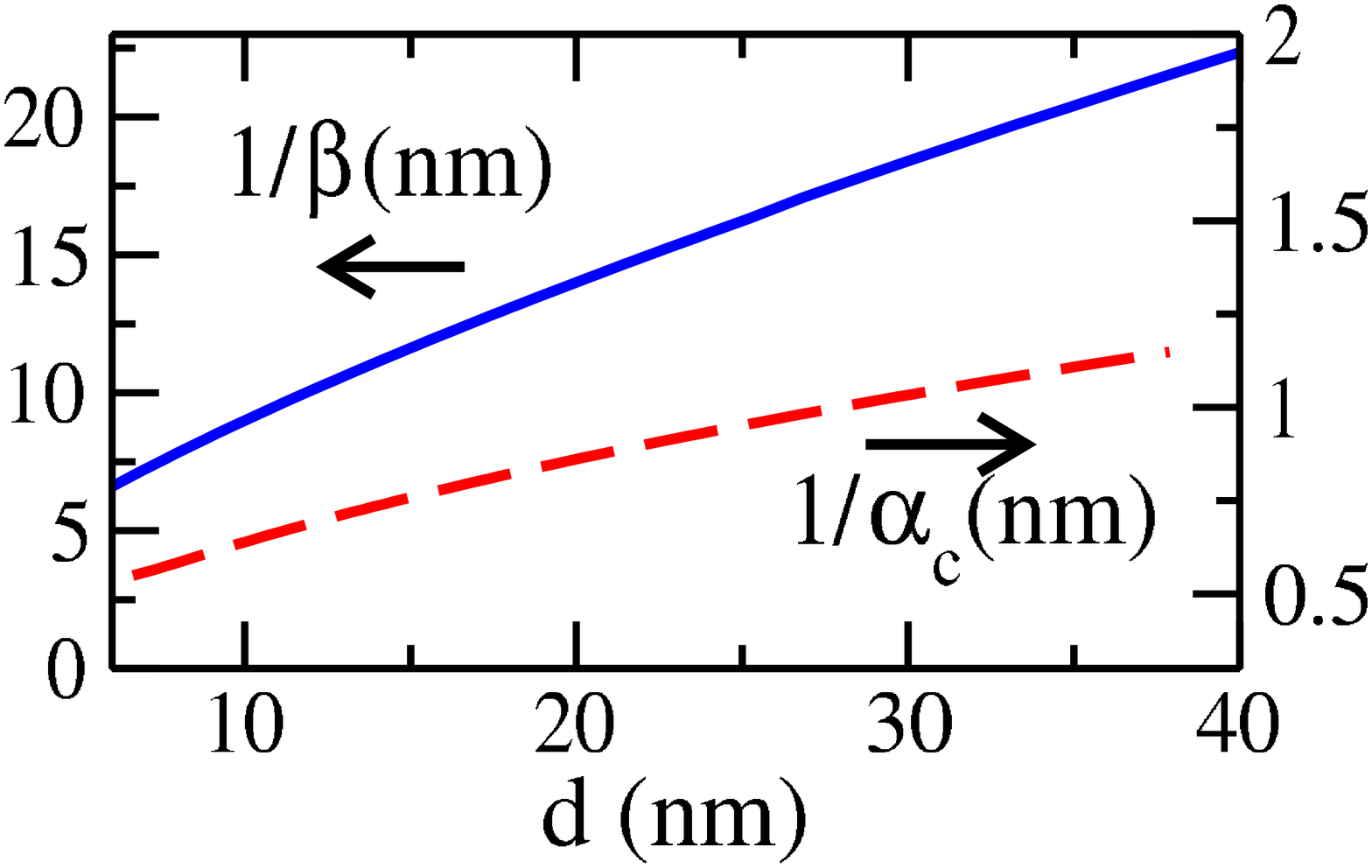}}
\caption{\label{fig:beta-alpha}(Color online)  Typical confinement lengths of the interface state in the $xy$-plane $1/\beta$ and $z$-direction $1/\alpha$. $\alpha$ depends on the value of the electric field applied: $\alpha_{\rm c}$ shown here corresponds to $F_{\rm c}(d)$.\\
\\}
\end{center}
\end{figure}

The inverse of the variational parameters, $1/\alpha$ and $1/\beta$,
are proportional to the confinement lengths in the $z$-direction and
the $xy$-plane respectively. Both depend on the distance $d$ of the
donor from the interface as shown in Fig.~\ref{fig:beta-alpha}, and
$1/\alpha$ also depends on the value of the
applied electric field, being larger for smaller $F$.
Fig.~\ref{fig:beta-alpha} gives the value of $1/\alpha$ for the electric
field $F_{\rm c}(d)$ at which the expectation values of the energies
of the states at the interface and at the donor are degenerate. The
expectation value of the position of the electron in the $z$-direction
is $\int_{-d}^{\infty} (z+d) f^2(z) \,\, dz = 5/\alpha$ from the
interface. This value is small compared to $d$ (for $d=30$~nm,
$5/\alpha= 5$~nm), justifying the validity of the two-well approach we
are using to solve the Hamiltonian.

\begin{figure}
\begin{center}
\resizebox{70mm}{!}{\includegraphics{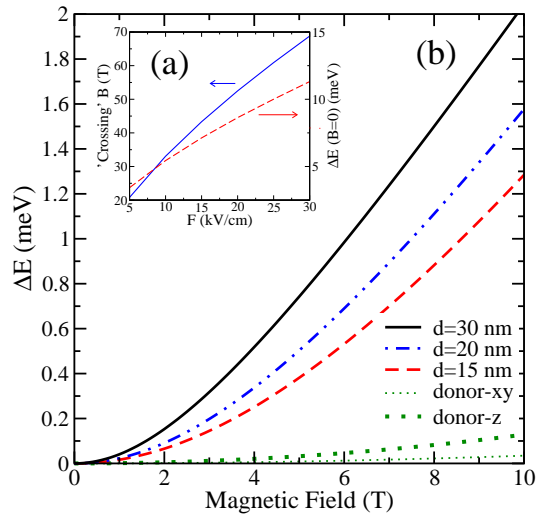}}
\caption{\label{fig:energy-B}(Color online) (a) (Right axis) Splitting
  between the $z$- and the $xy$-levels at the interface due to the
  anisotropic mass. (Left axis) Magnetic field at which the $xy$-levels
  become the ground state at the interface. (b) Energy shift of the
  interface ($z$-valleys) and donor levels (both $z$- and $xy$-valleys)
  under an applied magnetic field. Due to the mass anisotropy, the
  shift with magnetic field for the $z$-valleys levels is larger than
  for the $xy$-levels. At the donor, this implies that the degeneracy
  of the valleys is broken. At the interface, the $z$-valleys are the
  ground state up to very large values of the magnetic field as shown
  in (a).  }
\end{center}
\end{figure}

At the interface, the $z$-valleys' energy is lower than the
$xy$-valleys'. This is straightforward to show for an infinite
triangular potential~\cite{stern72} in which
$E_{z}(m_{\perp})/E_{z}(m_{\|})= \gamma^{1/3}=1.68637$ [see
Eq.~(\ref{eq:Etriangle})] . The difference between the levels depends on
the electric field as $F^{2/3}$. $\Delta
E=E_{z}(m_{\perp})-E_{z}(m_{\|})$ is shown in  Fig.~\ref{fig:energy-B}
(a) (right axis). For a field $F=5$ kV/cm, which is small for our
interests, the splitting is $\Delta E=2.5$~meV which corresponds to
$T\sim 30$ K. If a magnetic field is applied, the $z$-levels increase
their energy faster than the $xy$-levels until they cross. However, this
crossing happens at a very large magnetic field $B> 20$T (shown in the
left axis of Fig.~\ref{fig:energy-B} (a) as a function of the electric
field $F$). Therefore, for the range of parameters of interest here,
the $z$-valleys are always the ground state at the interface.
We point out that it is experimentally established that, in MOSFET
geometries equivalent to the one studied here, the interface ground
state is non-degenerate, with a 0.1 meV gap from the first excited
state.\cite{ando82} This is well above the operation temperatures in
the quantum control experiments investigated in the present context.

\begin{figure}
\begin{center}
\resizebox{40mm}{!}{\includegraphics{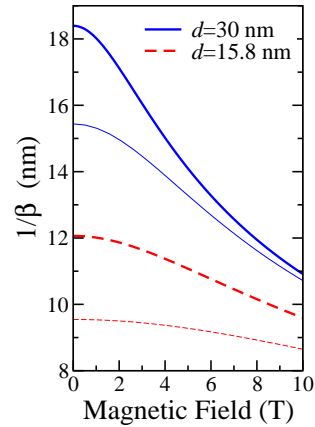}}
\caption{\label{fig:beta-overlap-versusB}(Color online) $1/\beta$
  versus magnetic field for two different interface-donor
  distances. Thick lines represent the variational results and the
  thin lines correspond to the harmonic approximation in
  Eq.~(\ref{eq:beta-B}).
}
\end{center}
\end{figure}

The magnetic field has two main effects on the states: (i) the
electron gets more confined in the direction parallel to the interface
and, consequently, (ii) its kinetic energy increases. The effect of
the magnetic field is strongest for the less confined wave-functions,
which correspond to the larger $d$'s. We can quantify the strength of
this effect by calculating  the magnetic field $B_{\rm c}$ that is
needed to get a magnetic length $\lambda_B$ of the same order of the
confinement length in the plane parallel to the interface: for a donor
a distance $d=30$~nm, $1/\beta=18.5$~nm and $B_{\rm c}\sim 2$ T while
for $d=15.8$~nm, $1/\beta=12$~nm and $B_{\rm c} \sim 4.5$ T. The
confinement effect is illustrated in
Fig.~\ref{fig:beta-overlap-versusB} where $1/\beta_B$ as a function of
magnetic field for two different values of $d$ is shown. The thick
lines correspond to the variational solution of  minimizing
\begin{equation}
H_{\rho} =- {{\partial^2}\over{\partial x^2}} - {{\partial^2}\over{\partial
 y^2}}+ {{1}\over{4}} \mu^2 \rho^2 +{{2Q}\over{\sqrt{\rho^2+d^2}}}
\end{equation}
with trial function $\sim \exp(-\beta_B^2 \rho^2/2)$. Closer
donors produce a larger confinement of the interface electron
wave-function but the effect of the magnetic field is much more
dramatic for the donors further away from the interface: for
$d=30$~nm, a magnetic field of $10$ T decreases the wave-function
radius by a $40\%$. Within the parabolic approximation  for the
interface potential $V_{\rm parab}(\rho)$ (which overestimates the
wave-function confinement) the dependence of $\beta_B$ on the
field is given by~\cite{qd-book}
\begin{equation}
\beta_B= \left(\beta^4+ 1/4\lambda_B^4\right)^{1/4}\,,
\label{eq:beta-B}
\end{equation}
with values as shown in Fig.~\ref{fig:beta-overlap-versusB} (thin lines).

The increase in energy with magnetic field, as calculated
variationally, is shown in Fig.~\ref{fig:energy-B} (b). We observe
again that the effect of the magnetic field is much stronger for the
larger values of $d$. The much smaller shift in the donor ground state
energy (discussed in the next subsection) is also shown for
comparison.


\subsection{Donor state $\psi_D$}
\label{subsec:donor}
The potential $V_D$ consists of the isolated impurity Coulomb potential
\begin{equation}
V_D=-{{2}\over{r}}.
\end{equation}

The solution of $H_D=T+V_D$ is taken to be of the form of the
anisotropic envelope wave-function,~\cite{Kohn} multiplied by $(z+d)$
to satisfy the boundary condition at the interface $\psi_D|_{z=-d}=0$
\begin{equation}
\psi_D=N
(z+d) e^{-\sqrt{{{\rho^2}/{a^2}}+{{z^2}/{b^2}}}} ,
\label{eq:psiD}
\end{equation}
where
$1/N^2=\pi a^2 b \left(d^2+b^2-{{1}\over{2}} b e^{-2d/b} \left(
{{1}\over{2}} d+b \right)\right)$, for $z \ge -d$. $\psi_D=0$ for $z
< -d$. For $d>> a, b$, $\psi_D$ reduces to the bulk limit $\psi_D^{\rm
  bulk}=1/(\pi a^2 b) \,
\exp{\left(-\sqrt{{{\rho^2}/{a^2}}+{{z^2}/{b^2}}}\right)}$. $a$ and
$b$ are variational parameters chosen to minimize the ground state
energy. Except for the smallest distances $d<2 a^* \sim 6$~nm, not
relevant here,  $a$ and $b$ coincide with the Kohn-Luttinger
variational Bohr radii of the isolated impurity ($d \rightarrow \infty
$) $a=2.365$~nm and $b=1.36$~nm.

\begin{figure}
\begin{center}
\resizebox{70mm}{!}{\includegraphics{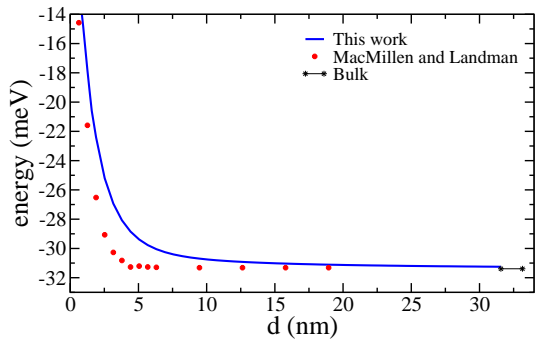}}
\caption{\label{fig:comp-MacMillen-Landman}(Color online)
Variational donor ground state energy $\langle \psi_D |H_D|
\psi_D \rangle$ versus $d$ obtained from our trial function $\psi_D$
and from the trial function used by MacMillen and
Landman~\cite{macmillen84} (data taken from Table III in
Ref.~\onlinecite{macmillen84}).
The short line at large $d$ represents the value of the
ground state  energy of an isolated donor in bulk Si (within the single valley
approximation) $\sim -31.2$~meV.~\cite{Kohn} A
``perfectly imaging'' plane is used ($\epsilon_{\rm oxide}=0$), as
assumed in Ref.~\onlinecite{macmillen84}.
For the intermediate and large values of $d$ of interest here,
results from the two approaches are in reasonable agreement.
}
\end{center}
\end{figure}

In Fig.~\ref{fig:comp-MacMillen-Landman} we show the
variational results for the ground state energy obtained from
our trial function $\psi_D$. For comparison, we also give results obtained
through the trial function  proposed by MacMillen and
Landman,\cite{macmillen84} where,
aiming at a good description for donors at
very short distances from the interface (typically smaller than the effective
Bohr radius $a^*$), a much larger set of variational states was used
for the expansion of the donor state. For this comparison, our results in
Fig.~\ref{fig:comp-MacMillen-Landman} correspond to a
``perfectly imaging'' plane ($\epsilon_{oxide}=0$), as assumed in
Ref.~\onlinecite{macmillen84}.
The energy depends strongly on $d$ for the smaller values of $d$,
and tends to the bulk value at long distances.
For the intermediate and large values of $d$ of interest here,
the two approaches are essentially equivalent.

We find that the effect of the external fields on the donor state is
negligible. For instance, for the largest electric fields of interest
here ($F_{\rm c}$ at short distances $d \sim 6$~nm), the energy
corresponding to the electric field potential is $\langle \psi_D|
\kappa eFz |\psi_D \rangle =0.18 \, Ry^*$, to compare with
$1.6 \,Ry^* $ for the isolated donor ground state in bulk. The effect
of a magnetic field on the electron wave-function at the donor is also
very small: the field required to get a magnetic length of the order
of the Bohr radius $a=2.365$~nm is $B_{\rm c}\sim 120$ T! The donor
ground state energy shift due to the magnetic field can be estimated
by\cite{landauQM}
\begin{equation}
\Delta E= {{\langle \tilde{r}^2 \rangle} \over {4 \lambda_B^4}} Ry^* \,\, ,
\end{equation}
where lengths are given in units of $a^*$,  $\langle \tilde{r}^2
\rangle = 2a^2/a^{*2}$ for the $z$-envelopes, and $\langle \tilde{r}^2
\rangle = (a^2+b^2)/a^{*2}$ for the $xy$-envelopes. The results are
shown in Fig.~\ref{fig:energy-B}(b) and are comparable to the values
for the 1s orbital of shallow donors calculated numerically in
Ref.~\onlinecite{thilderkvist94}. Note that the magnetic field
partially  breaks the six-valley degeneracy due to the different
confinement radius of the electron wave-function in each of the
different valleys.

\begin{figure}
\begin{center}
\resizebox{70mm}{!}{\includegraphics{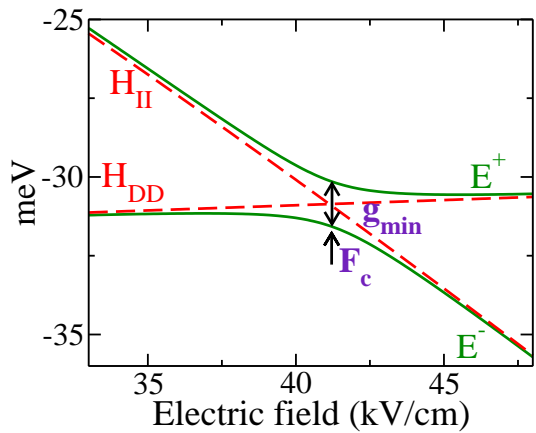}}
\caption{\label{fig:anticrossing}(Color online) Eigenenergies $E^+$ and $E^-$
  of the Hamiltonian in Eq.~(\ref{H}) for $d=11$~nm. They
  show an anticrossing behavior with a minimum gap $g_{\rm min}$ at
  $F_{\rm c}$. Tunneling times are related to $g_{\rm min}$ as
  $\tau=\hbar/g_{\rm min}$. $F_{\rm c}$ can be determined as the field at
  which $H_{II}=H_{DD}$.}
\end{center}
\end{figure}

\subsection{Shuttling between interface and donor states}
\label{subsec:shuttling}
\begin{figure}
\begin{center}
\resizebox{80mm}{!}{\includegraphics{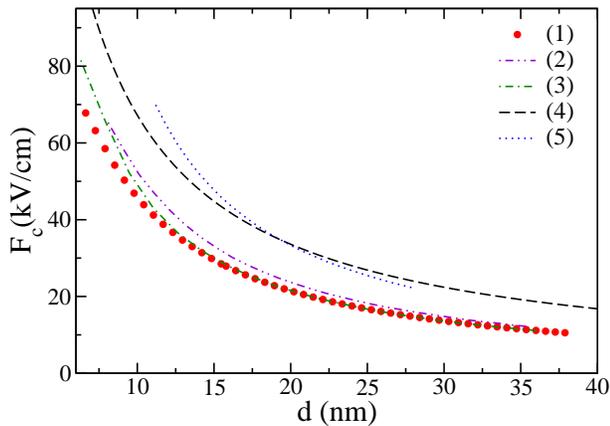}}
\caption{(Color online) Comparison of $F_{\rm c}$ versus $d$ for the
  single donor
 problem (P) obtained from different forms for $\psi_D$. Label (1)
 corresponds to
  $\psi_D$ with anisotropic mass as defined in
  Eq.(\ref{eq:psiD}). Label (2) assumes isotropic effective
mass $m^*=0.29819 m$,
  chosen so that the ground state energy is $-31.2$~meV, the same as
  for the single valley approximation. Label (3) corresponds to
  $\psi_D \sim
  \exp{\left(-\sqrt{{{\rho^2}/{a^2}}+{{z^2}/{b^2}}}\right)}$, and the
  same ground state energy. Label (4) reproduces the tight-binding
  results in Ref.~\onlinecite{martins04}, where the six valley
  degeneracy is considered, leading to the ground state energy of
  $-45$~meV. The latter coincides with the experimental binding energy
  for a P donor in Si. Label (5) considers $\psi_D \sim \exp(-r/a)$
  with isotropic mass $m^*=0.43 m$ so that the ground state energy is
  $-45$~meV. \\
}
\label{fig:all-approaches-P}
\end{center}
\end{figure}

We model the donor electron ionization under an applied electric field
along $z$ by considering the tunneling process from the donor well into the
triangular well at the Si/SiO$_2$ interface (see Fig.~\ref{fig:potentialF}).
The required value of the field for ionization to take place may be
estimated from $H_{II}=H_{DD}$ [see Eq.~(\ref{eq:2by2})]. We call  $F_{\rm c}$ the characteristic field
for which this condition is fulfilled, which is equivalent to require
that the gap between the two eigenenergies $E^+$ and $E^-$ of $H$ is
minimum, as illustrated in Fig.~\ref{fig:anticrossing}.

Our results for $F_{\rm c}$ versus $d$ are shown by the solid dots
[labelled (1)]  in Fig.~\ref{fig:all-approaches-P}.
In this figure we also test the robustness of our approach, namely
using $\psi_D$ as given in Eq.~(\ref{eq:psiD}),
by comparing the values of $F_{\rm c}$ obtained assuming different forms
for the donor trial function.
Curves (2) and (3) correspond to isotropic and anisotropic
wave-functions respectively, with the same ground state
energy for the electron at the donor as obtained from  $\psi_D$ in
Eq.~(\ref{eq:psiD}), $\sim -31.2$~meV. Note that they compare
very well with curve (1). Curve (4) corresponds to a tight-binding
result~\cite{martins04} in which the six-valley degeneracy of Si is
incorporated. Although all curves are qualitatively similar,
curve (4) is shifted towards larger fields. The origin of this shift
is investigated by considering an isotropic trial function whose parameters
have been chosen to give a ground state energy $\sim -45$~meV, and we
note that the results, given in curve (5), compare very well with
those in curve (4). We conclude that the shift in the value of $F_{\rm
  c}$ when the six-fold degeneracy of the Si conduction band is
considered is mainly due to the fact that the ground state energy at
the donor in the single valley approximation [curve (1)] is $\sim
-31.2$~meV while it is $\sim -45$~meV when the intervalley coupling is
included [curve (4)]. In the following we use $\psi_D$ as defined in
Eq.~(\ref{eq:psiD}), keeping in mind that electric field values are
bound to be somewhat underestimated.

\begin{figure}
\begin{center}
\resizebox{80mm}{!}{\includegraphics{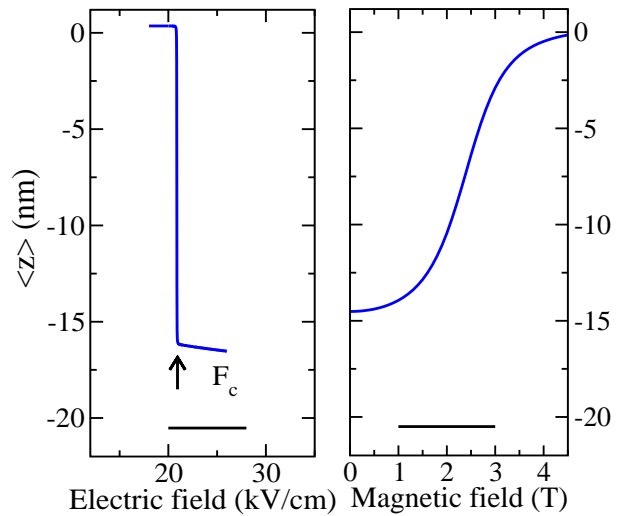}}
\caption{\label{fig:position}(Color online)   Expectation value of the electron $z$-coordinate $\langle z \rangle =\langle \Psi^-|z|\Psi^-\rangle$  for $d=20.5$~nm versus (a) electric field and (b) magnetic field [$F= F_{\rm c} +60 $ V/cm]. The horizontal lines represent the  relative position of the interface. The donor is at $\langle z \rangle = 0$. The electric field moves the electron from the donor to the interface while a parallel magnetic field takes the electron back to the donor.}
\end{center}
\end{figure}

We may picture the electron shuttling between the two wells under
applied electric and magnetic fields by calculating the
expectation value of its position along $z$ at the ground state,
$\langle z \rangle=\langle \Psi^- | z| \Psi^- \rangle$. The
results for $d=20.5$~nm  are shown in Fig.~\ref{fig:position} ,
where the horizontal lines mark the position of the interface. The
distance between $\langle z \rangle$ and the interface tends to
$5/\alpha$ for $F \geq F_{\rm c}$, where $\alpha$ also depends on
$F$. In Fig.~\ref{fig:position}(a) we show pictorially how the
electron would evolve from the donor to the interface well when an
electric field is applied. At small values of $F$ the electron is
at the donor well, $\langle z \rangle \sim 0$ and $\Psi^- \approx
\psi_D$. The center of mass is slightly shifted from the donor
site due to the factor $(z+d)$ in $\psi_D$. Above $F_{\rm c}$, the
electron eventually tunnels to the interface. Starting with the
electron at the interface in a near-degeneracy configuration ($F
\gtrsim F_{\rm c}$), a relatively modest magnetic field can cause
the electron to move in a direction {\em parallel} to the field
and against the electric field, as shown in
Fig.~\ref{fig:position}(b). This is due to the much larger shift
of the interface state energy with magnetic field compared to the
shift of the donor ground state energy [see
Fig.~\ref{fig:energy-B}(b)]. This behavior characterizes electrons
originating from the donors, and not other charges that the SET
may detect, like charges in metallic grains on the device
surface.~\cite{kenton06} The combination of parallel electric and
magnetic fields constitutes therefore a valuable experimental
setup to investigate whether charge detected at the interface
actually originates from a donor.~\cite{CKDmag}

\begin{figure}
\begin{center}
\resizebox{70mm}{!}{\includegraphics{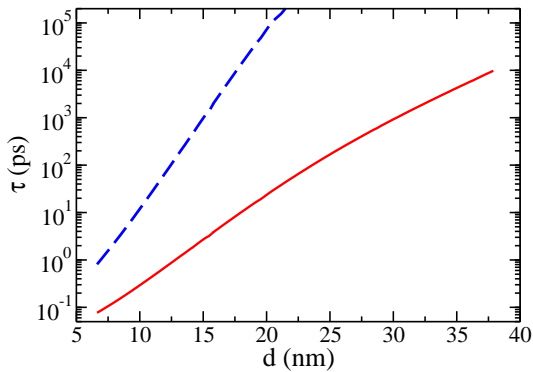}}
\caption{\label{fig:times}(Color online) Donor ionization tunneling
  (solid line) and adiabatic passage (dashed line) times versus $d$.
}
\end{center}
\end{figure}

A key parameter determining the feasibility of quantum computation in
the doped Si architecture is the time required to shuttle the electron
between the donor and the interface. This time should be orders of
magnitude smaller than the coherence time to allow for many
operations and error correction while coherent evolution
of the qubit takes place.
The tunneling process conserves the spin, but  coherence would be lost
for orbital/charge degrees of freedom. Therefore, if
quantum information is stored in a charge qubit, the electron should
evolve adiabatically from the donor to the interface, while tunneling
would be acceptable for spin qubits.
In an adiabatic process the modification of the Hamiltonian (for
instance, when an external field is applied) is slow enough that the
system is always in a known energy eigenstate, going continuously
from the initial to the final eigenstate.~\cite{messiah} Here, we
calculate both tunneling and the adiabatic passage times.

We estimate the tunneling time from the minimum gap $g_{\rm min}$
between the two eigenvalues $E^+$ and $E^-$ (see
Fig.~\ref{fig:anticrossing}) via the uncertainty relation
$\tau=\hbar/g_{\rm min}$. The adiabatic time is calculated
as~\cite{ribeiro02,martins04} $\tau_a = \hbar |e| F_{\rm max} d/ g_{\rm min}^2$
and is orders of magnitude larger than the tunneling time. $F_{\rm
  max}$ is chosen so that the electron is at the interface $\Psi^-
\approx \psi_I$. The results for the tunneling and adiabatic passage
times (for $F_{\rm max}=2 F_{\rm c}$) are shown in
Fig.~\ref{fig:times}. The times depend exponentially on the distance
$d$. Tunneling times range from $0.1$ ps for $d=6$~nm to $10$~ns for
$d=38$~nm. Adiabatic times range from $1$ ps for $d=6$~nm to $100$~ns
for $d=20$~nm and get very large at longer distances. These times are
to be compared to the experimental values of spin coherence and
charge coherence respectively.

Spin dephasing in Si is mainly due to dipolar
fluctuations in the nuclear spins in the system, which produce a
random magnetic field at the donor electron spin. The
spin dephasing time in bulk natural Si is $T_2 \sim 1$
ms.~\cite{sousa03,abe04,witzel05,tyryshkin06} Natural Si is mostly composed
of $^{28}$Si (no nuclear spin), with a small fraction
($4.67 \%$) of  $^{29}$Si (nuclear spin $1/2$), therefore, $T_2$ can
be dramatically improved through isotope
purification~\cite{sousa03,abe04,witzel05,tyryshkin06} up to $100$ ms
or longer in bulk. Moreover, it has been recently proposed that the
spin dephasing times can be arbitrarily prolonged by applying designed
electron spin resonance pulse sequences.~\cite{witzel06} On the other
hand, the closeness of a surface or interface can reduce the spin dephasing
times.~\cite{schenkel06} The tunneling time calculated here
is orders of magnitude smaller than the bulk $T_2 \sim 1$ ms for
natural Si: In the worst case scenario (long distances $d \sim 40$~nm)
$T_2/\tau \sim 100$. Therefore, we expect tunneling times to be always
much shorter than decoherence times even in the presence of an
interface.

Charge coherence time is considerably shorter than spin coherence time, since
charge couples very strongly to the environment through
the long range Coulomb interaction. The main channels of decoherence
are charge fluctuations and electron-phonon
interactions.~\cite{Hayashi,hu05PRB} The charge coherence time has
been measured to be $\sim 200$ ns for Si quantum dots surrounded by oxide
layers.~\cite{gorman05} This number has to be compared to the
calculated adiabatic times, which are much longer than the tunneling
times. Therefore, for charge qubits to be realizable in the
configuration discussed here,  the donor-interface distance $d$ has to
be limited to a maximum of $20$~nm.

 \section{Donor pair}
\label{sec:donor-pair}
\subsection{Planar density}
\label{subsec:planar-density}
\begin{figure}
\begin{center}
\resizebox{80mm}{!}{\includegraphics{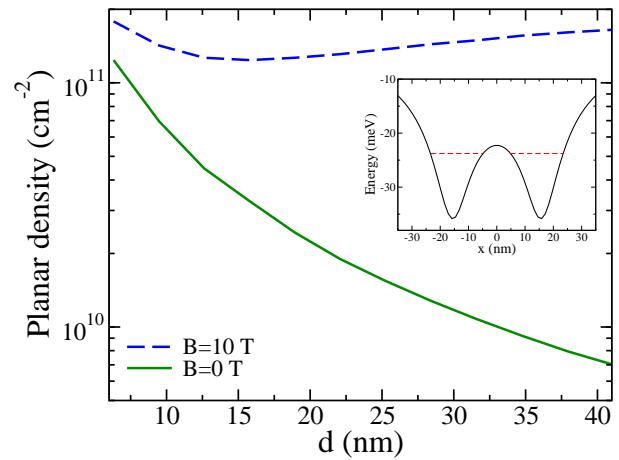}}
\caption{\label{fig:double-well}(Color online) The main panel gives
  the maximum donor planar density $n_{\rm max}$ for which electrons
  drawn to the interface remain localized around the donor region,
  and do not form a
  2DEG. It is assumed that all donors are at the same
  distance from the interface $d$. $n_{\rm max}$ is estimated
  from the criterion that the barrier between neighboring wells at
  the interface is above $E_{\rm DW}$ given in
  Eq.~(\ref{eq:Rmin-condition}).
  Results are shown for $B=0$ T and $B=10$ T. The inset shows the
  double well potential parallel to the interface created by two
  donors located a distance $d=6.3$~nm from the interface and separated by
  $R=28$~nm.  The expectation value of the energy $E_{\rm DW}$
  (given by the dashed lines for $B=0$) is also lower than the
  single-donor well ground state energy. }
\end{center}
\end{figure}

We estimate the maximum planar density of donors
($n_{\rm max}$) allowed to avoid the formation of a
2DEG as $n_{\rm max}=1/R_{\rm min}^2$, where $R_{\rm min}$ is a
minimum distance between two donors which is calculated as follows: We
assume two donors located at the same distance $d$ from the
interface and a distance $R$ apart. The resulting
double well potential along the interface $xy$-plane is
\begin{eqnarray}
V_{\rm DW} (x,y) &=& {2(Q-1)\over\sqrt{(x+ R/2)^2+y^2+d^2}}\nonumber\\ &+&
{2(Q-1)\over\sqrt{(x- R/2)^2+y^2+d^2}}~,
\end{eqnarray}
as illustrated in the inset of Fig.~\ref{fig:double-well} for
$d=6.3$~nm and $R=28$~nm.
We adopt two different criteria to estimate $R_{\rm min}$. (i) The first one
requires $R_{\rm min}=2/\beta$ where
$2/\beta$ is the width of the gaussian $g(\rho)$  (see
Fig.~\ref{fig:beta-alpha}). For instance, for $d= 30$~nm, this gives
$R_{\rm min} \approx 40$~nm leading to
$n_{\rm max} \approx 6 \times 10^{10}$ cm$^{-2}$, while for $d=10$ nm,
$R_{\rm min} \approx 18$ nm and $n_{\rm max} \approx  3 \times
10^{11}$ cm$^{-2}$. (ii) The second criterion,
which we find to be slightly more restrictive, requires a high enough barrier within the double
well, and is given by
\begin{equation}
E_{\rm DW}= \langle \psi_I |\, H_{\rm DW} \,| \psi_I \rangle \le
V_{\rm DW}(0,0)\,,
\label{eq:Rmin-condition}
\end{equation}
where $R_{\rm min}$ corresponds to the equality condition. $H_{\rm DW}$ is the double-well  2-dimensional Hamiltonian
\begin{equation}
 H_{\rm DW}= T_x+T_y+V_{\rm DW}(x,y) \,,
\end{equation}
with $T_x$ and $T_y$ the kinetic energy terms.
$V_{\rm DW}(0,0)$ is the maximum height of the inter-well barrier, which is
required to be above the single-particle expectation value of the
energy $E_{\rm DW}$.
The maximum  planar density
estimated from Eq.~(\ref{eq:Rmin-condition})
is shown in the main panel of Fig.~\ref{fig:double-well}.
For instance,  $n_{\rm max}(d=30 \,{\rm nm}) \approx
10^{10}$~cm$^{-2}$ obtained from $R_{\rm min} \approx 88$~nm.
$n_{\rm max}$ is larger for the donors closer to the interface. For
instance, $n_{\rm max}(d=10 \,{\rm nm}) \approx 6\times
10^{10}$ cm$^{-2}$  ($R_{\rm min} \approx 38$~nm).

As shown in Fig.~\ref{fig:beta-overlap-versusB},
$1/\beta_B$ decreases with a perpendicular magnetic field, hence
increasing the maximum planar density. The first criterion for the
maximum planar density gives, for $d=30$~nm and $B=10$~T, $R_{\rm
  min}\approx 22$~nm and $n_{\rm max} \approx  2\times
10^{11}$~cm$^{-2}$. The second criterion gives the dashed curve in
Fig.~\ref{fig:double-well} which is, on average,
almost one order of magnitude larger than without a magnetic
field. Note that the effect of the magnetic field is much stronger for
large distances $d$.

\subsection{Qubit interaction at the interface: exchange}
\label{subsec:exchange}
One of the problems for quantum computation in doped Si arises from
the lack of
control of the exact position of the donors. The main consequence of
this is the indetermination of the value of the exchange between two
neighboring donor electrons due to the theoretically predicted
oscillations  of exchange with $R$, caused by valley interference
effects.~\cite{KHD1,KHD2}  One straightforward way to alleviate this
problem is to perform these operations at the interface~\cite{CKDmag}
where, as discussed in Sec.~\ref{subsec:interface}, this degeneracy is
lowered. Additionally, it would be much easier to control
the qubit operations when the electrons are at the interface,
similar to the successful experiments on double quantum dots in
GaAs~\cite{petta05,koppens05,johnson05,laird06} and Si.~\cite{gorman05}  Note
that the potential created by donor pairs (inset in
Fig.~\ref{fig:double-well}) resembles very much a double quantum dot
with the clear advantage that, in this case, the potential is produced
exclusively by the Coulomb attraction of the donors and its exact form
is known: $V = 2(Q-1)/\sqrt{\rho^2+d^2}$.

It is therefore of interest to determine the exchange coupling between
donor electrons at the interface. As a first approach, we perform
these calculations within the Heitler-London method. The validity and
limitations of this method to calculate exchange in semiconductor
nanostructures has been previously discussed by the
authors.~\cite{CKDexch} The expression for the exchange within
this approximation is
\begin{eqnarray}
J&=&{{2 S^2}\over{1-S^4}} \langle \Phi_L(1) \Phi_R(2) | 2 H_{\rm DW}(1)+ {{e^2}\over{\epsilon_{\rm Si} r_{12}}}|\Phi_L(1) \Phi_R(2) \rangle  \nonumber \\
&-& {{2}\over{1-S^4}} \langle \Phi_L(1) \Phi_R(2)| 2 H_{\rm
DW}(1)+ {{e^2}\over{\epsilon_{\rm Si} r_{12}}} |\Phi_L(2)
\Phi_R(1) \rangle
\nonumber\\
\label{eq:exch}
\end{eqnarray}
where $\Phi_{L,R}= g(x \mp R/2,y)$, $S=\exp[-\beta^2 (R/2)^2]$ is
the overlap, and $e^2/\epsilon_{\rm Si} r_{12}$ is the
electron-electron interaction with $r_{12}$ the distance between
electron (1) and electron (2). The first term in
Eq.~(\ref{eq:exch}) is the direct term and the second is the
exchange term.

\begin{figure}
\begin{center}
\resizebox{70mm}{!}{\includegraphics{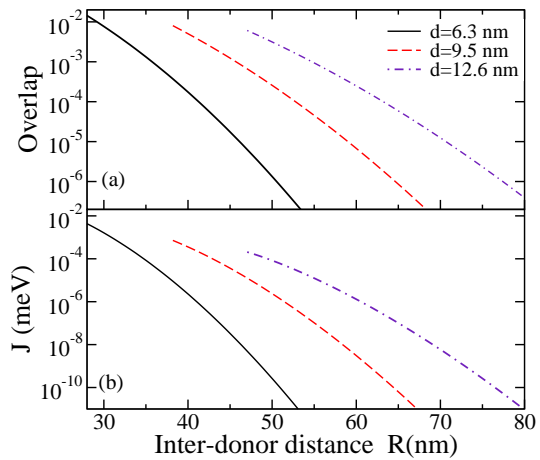}}
\caption{\label{fig:exchange-at-interface-R}(Color online) (a) Overlap
  $S$ between electron wavefunctions at neighboring wells at the
  interface for three different values of $d$. (b) Exchange $J$
  calculated within the Heitler-London approximation.
  Values for $S$ and $J$ given here satisfy $n<n_{\rm max}$.}
\end{center}
\end{figure}

In Fig.~\ref{fig:exchange-at-interface-R} we show the exchange $J$ and
the overlap $S$ versus the inter-donor distance $R$ for three
different values of $d$.  Note that the overall dependence of these
two quantities with $R$ is very similar, indicating that the behavior
of $J$ is closely related to the  overlap.\cite{CKDexch}
$J$ and $S$ values are shown only for
distances $R >R_{\rm min}$, defined in
Sec.~\ref{subsec:planar-density}. For a wide range of $R$'s, $J$ is of
the same order as in GaAs double quantum dots where $J$'s as low as
$\sim 10$~neV have been measured.~\cite{laird06}

\begin{figure}
\begin{center}
\resizebox{60mm}{!}{\includegraphics{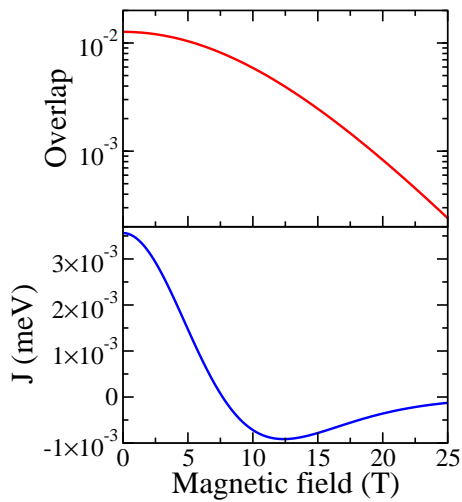}}
\caption{\label{fig:exchange-at-interface-mag}(Color online)
  Modulation of overlap and exchange by a magnetic field perpendicular
  to the interface in the
  particular case $d=6.3$~nm and $R=28$~nm. The
  double well potential for these
  particular values of $d$ and $R$ is depicted in the inset of
  Fig.~\ref{fig:double-well}.}
\end{center}
\end{figure}

Exchange control can be performed by applying a
magnetic field perpendicular to the interface, which
reduces the wave-function radius, therefore decreasing the
overlap. The effect of the magnetic field on the exchange is well
known:~\cite{Exch} $J$ decreases and eventually changes sign when the
triplet is favored becoming the ground state.
This is illustrated in Fig.~\ref{fig:exchange-at-interface-mag} for
$d=6.3$~nm and $R=28$~nm.
At very large fields the singlet and triplet states become degenerate,
as expected. Note that, in this case, the qualitative behavior of $S$
and $J$ with magnetic field is very different.

\section{Summary and conclusions}
\label{sec:discussion}
Quantum computer architectures based on semiconductor nanostructure qubits have the key potential advantage of scalability, which has led to the great deal of current interest in Si- and GaAs-based quantum computation. Silicon based spin qubits have the important additional advantage of extremely long spin coherence times ($T_2^{\rm Si} \sim$ miliseconds or more) since isotopic purification (eliminating $^{29}$Si nuclei) could considerable suppress spectral diffusion induced electron spin decoherence~\cite{sousa03,tyryshkin03,abe04,witzel05,tyryshkin06,witzel06} leading to $T_2^{\rm Si} \sim 100$ ms, whereas electron spin coherence time is constrained to be rather short in GaAs quantum dot structures, $T_2^{\rm GaAs} \sim 1-10 \,\mu$s, since neither Ga nor As nuclei have zero nuclear angular momentum isotopes. However, the experimental progress in Si spin qubits has been very slow whereas there has been impressive recent experimental progress in the GaAs quantum dot spin qubits.~\cite{petta05,koppens05,johnson05,koppens06,laird06} The main experimental advantage of GaAs quantum dot system has been the ease in the 1-qubit and 2-qubit manipulation because the electrons near the surface can be effectively controlled by surface gates. By contrast, Si:P qubits are in the bulk, severely hindering experimental progress since electron manipulation in the bulk has turned out to be a difficult task.

In this paper we show through detailed quantitative theoretical work how to control and manipulate qubits (i.e. both single electrons and two-electron exchange coupling) in a doped-Si quantum computing architecture~\cite{Kane} by applying external electric and magnetic fields. In particular, we have analyzed three main issues: (i) the times involved in the donor electron 'shuttling' between the donor and the interface of Si with (typically) SiO$_2$ have been found to be a few orders of
magnitude shorter than the spin coherence times in Si, as required to allow for the necessary 'logic operations' and 'error correction' to take place; (ii) the
existence of a well defined interface state where the electron remains
bound and localized, so that it does not spread and form a 2DEG.
This condition, which guarantees that electrons actually involved in a
particular operation be taken back from the interface to donor sites,
leads to a lower bound for the interdonor spacing, and consequently a maximum
donor planar density; and (iii) the possibility of performing the
two-qubit exchange gate operations at the interface, instead of around
the donor sites as originally proposed.\cite{Kane} Our results show that sufficiently large values of exchange coupling ($\sim 10^{-4}$ meV) can be achieved.

Interface operations have several potential advantages over bulk
operations, the most
obvious one being that the read-out procedure would be simplified.
A well known limitation of exchange gates for donor electrons in the
bulk is the
oscillatory behavior of the exchange coupling, which is due to the
strong pinning of the six conduction band Bloch-function phases at
each donor site, where the Coulomb potential is
infinitely attractive.\cite{KCHD} This condition is alleviated at the interface in two ways: first, the six valley degeneracy is partially lifted and, second,
although the electrons at the interface remain bound to the donors,
the binding potential is not singular;
it is actually equivalent to a quantum dot potential. Experiments on charge-qubit control in a double quantum dot at Si surface~\cite{gorman05} indicate that the exchange
oscillatory behavior may not be a severe problem for donor-bound
electrons manipulated at the Si/SiO$_2$ interface.

Our proposal combines the advantages of Si spin qubits (i.e. long $T_2$ time) with the structural advantages of GaAs qubit control and manipulation. We believe that the specific experiments we propose (and analyze in quantitative details) in this paper, if carried out, will go a long way in establishing the feasibility of a Si quantum computer.

%\begin{figure}
%\begin{center}
%\resizebox{80mm}{!}{\includegraphics{}}
%\caption{\label{fig:}(Color online) }
%\end{center}
%\end{figure}
%

\begin{acknowledgments}
This work is supported by LPS and NSA. B.K. also acknowledges support from CNPq, FUJB, Millenium Institute on Nanotechnology - MCT, and FAPERJ.

\end{acknowledgments}

\bibliography{long}

\end{document}